\documentclass[%
aip,
 amsmath,amssymb,
reprint,
]{revtex4-1}

\usepackage{graphicx}% Include figure files
\usepackage{dcolumn}% Align table columns on decimal point
\usepackage{bm}% bold math
\usepackage[utf8]{inputenc}
\usepackage[T1]{fontenc}
\usepackage{mathptmx}
\usepackage{etoolbox}

\usepackage{amsmath}
\usepackage{tabularx}
\usepackage{comment}
\usepackage{placeins}

\usepackage{longtable}

\usepackage{supertabular}
\usepackage{booktabs}
\usepackage{multirow}

\newcolumntype{R}{>{\RaggedRight\let\newline\\\arraybackslash\hspace{0pt}}X}

\makeatletter
\def\@email#1#2{%
 \endgroup
 \patchcmd{\titleblock@produce}
  {\frontmatter@RRAPformat}
  {\frontmatter@RRAPformat{\produce@RRAP{*#1\href{mailto:#2}{#2}}}\frontmatter@RRAPformat}
  {}{}
}%
\makeatother

\begin{document}

\preprint{AIP/123-QED}

\title[Spectroscopic characterization of singlet-triplet doorway states of aluminum monofluoride]{Spectroscopic characterization of singlet-triplet doorway states of aluminum monofluoride}

% Force line breaks with \\
\author{N. Walter}
 \homepage{https://www.fhi.mpg.de/207024/mp-department}
 \email{walter@fhi-berlin.mpg.de}
 \author{J. Seifert}
  \author{S. Truppe}
 \author{H.C. Schewe}
 \affiliation{Fritz-Haber-Institut der Max-Planck-Gesellschaft, Faradayweg 4-6, 14195 Berlin, Germany}
 
\author{B.G. Sartakov}
 \affiliation{Prokhorov General Physics Institute, Russian Academy of Sciences, Vavilovstreet 38, 119991 Moscow, Russia}
\author{G. Meijer}
 \email{meijer@fhi-berlin.mpg.de}
 \affiliation{Fritz-Haber-Institut der Max-Planck-Gesellschaft, Faradayweg 4-6, 14195 Berlin, Germany}

\date{\today}% It is always \today, today,

\begin{abstract}
%general
Aluminum monofluoride (AlF) possesses highly favorable properties for laser cooling, both via the A$^1\Pi$ and a$^3\Pi$ states. Determining efficient pathways between the singlet and the triplet manifold of electronic states will be advantageous for future experiments at ultralow temperatures. The lowest rotational levels of the A$^1\Pi, v=6$ and b$^3\Sigma^+, v=5$ states of AlF are nearly iso-energetic and interact via spin-orbit coupling. These levels thus have a strongly mixed spin-character and provide a singlet-triplet doorway. We here present a hyperfine resolved spectroscopic study of the A$^1\Pi, v=6$ // b$^3\Sigma^+, v=5$ perturbed system in a jet-cooled, pulsed molecular beam. 
From a fit to the observed energies of the hyperfine levels, the fine and hyperfine structure parameters of the coupled states, their relative energies as well as the spin-orbit interaction parameter are determined. The standard deviation of the fit is about 15~MHz.
We experimentally determine the radiative lifetimes of selected hyperfine levels by time-delayed ionization, Lamb dip spectroscopy and accurate measurements of the transition lineshapes. The measured lifetimes range between 2 ns and 200 ns, determined by the degree of singlet-triplet mixing for each level. 

\end{abstract}

\maketitle

\section{\label{sec:introduction}Introduction}

In 1939, Rochester recorded the first electronic absorption spectrum of aluminum monofluoride (AlF).\cite{rochester1938} He attributed the bands centered around 227 nm to the A$^1\Pi$ -- X$^1\Sigma^+$ system and concluded that the vibrational and rotational constants of these coupled electronic states are almost the same. By 1974, a total of nine singlet and seven triplet electronic states had been assigned to AlF and ordered into a common energy scheme.\cite{1974barrow} The energy of the triplet relative to the singlet manifold of states was determined from observed spectral perturbations, in particular those due to spin-orbit coupling between the A$^1\Pi$ and b$^3\Sigma^+$ states. The direct a$^3\Pi$ -- X$^1\Sigma^+$ singlet-triplet transition was only observed two years later. \cite{1976kopp,1976rosenwaks}

A few years ago, our group became interested in AlF in view of its predicted
favorable properties for laser cooling and magneto-optical trapping. \cite{dirosa,wells} We determined precise molecular parameters for the lowest singlet and triplet states, \cite{alf,max,nicoleastate} thereby establishing AlF as a new benchmark molecule for quantum chemistry calculations. We studied the spin-orbit interaction between the $v=0$ level in the A$^1\Pi$ state and the various vibrational levels in the b$^3\Sigma^+$ state to determine the fraction of triplet character of the A$^1\Pi, v=0$ state.\cite{max} This fraction is low enough, i.e. this particular interaction is weak enough, to not cause a significant loss (< 10$^{-6}$) for optical cycling on the A$^1\Pi, v=0$ -- X$^1\Sigma^+, v=0$ band.\cite{Simon} 

Although the perturbation of the A$^1\Pi, v=0$ state is weak, there is a strong interaction between the nearly iso-energetic lowest rotational levels of the A$^1\Pi, v=6$ and b$^3\Sigma^+, v=5$ states. \cite{1974barrow} This results in a mixed singlet-triplet character of these rotational levels that allows them to act as a path from the singlet to the triplet manifold. Such strongly mixed levels have been proposed as doorways, for instance, to produce translationally cold CO molecules in their absolute ground-state, starting from trapped molecules in the a$^3\Pi$ state.\cite{blokland}

\begin{figure}[b]
		\centering
		\includegraphics[width=0.48\textwidth,clip]{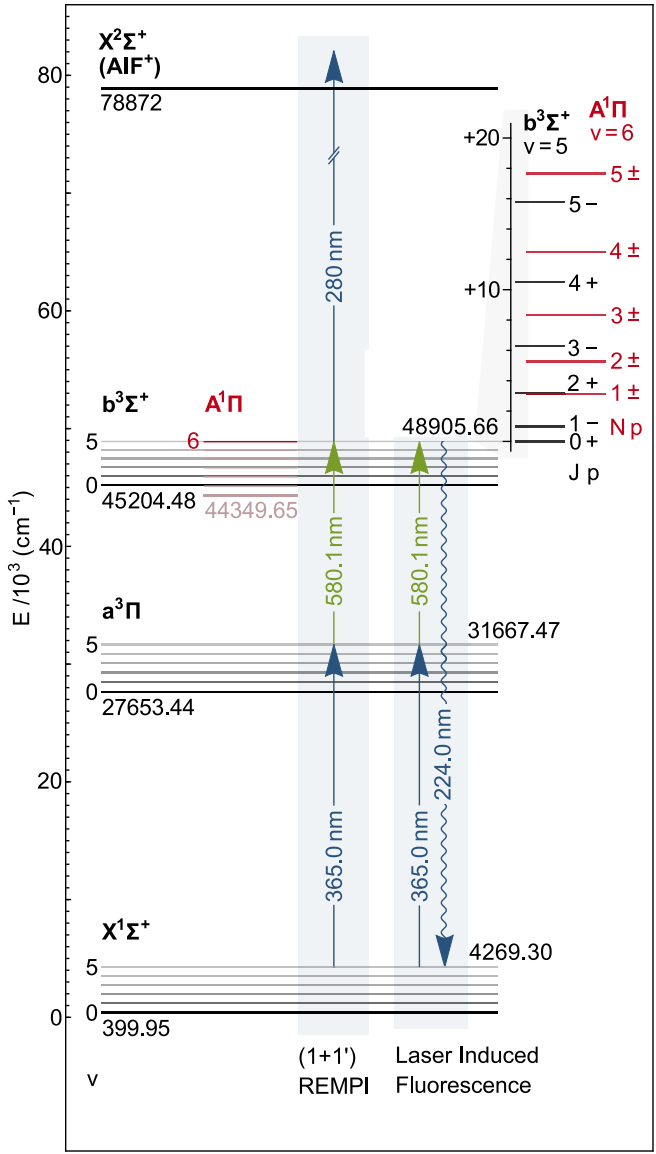}
	\caption{\label{fig:EnergyScheme} Scheme of the energy levels and of the excitation and detection methods relevant for this study. The top right part shows in detail the unperturbed rotational levels of the A$^1\Pi, v=6$ and b$^3\Sigma^+, v=5$ states.
	}
\end{figure}

Here, we present a detailed spectroscopic investigation of the lowest rotational levels of the perturbed  A$^1\Pi, v=6$ // b$^3\Sigma^+, v=5$ system of AlF. In Figure~\ref{fig:EnergyScheme} the relevant energy levels are shown together with the excitation and detection schemes that are used. All experiments are carried out on a jet-cooled supersonic molecular beam containing AlF, as briefly described in Section~\ref{sec:Setup}. Using two-color resonance enhanced multiphoton ionization ((1+1$'$)~REMPI), we record rotationally resolved spectra of the perturbed region, presented in Section~\ref{sec:Rot}. Subsequently, hyperfine resolved laser-induced fluorescence (LIF) spectra are recorded using a narrow-band continuous wave laser for A$^1\Pi, v=6$ // b$^3\Sigma^+, v=5 \leftarrow$ a$^3\Pi, v=5$ excitation (see Section~\ref{sec:Hyperfine}). The radiative lifetimes of the perturbed levels range between 2 ns and 200~ns and are determined from spectral line broadening, Lamb dip spectrosopy and time-delayed ionization, as described in Section~\ref{sec:Lifetimes}.

\section{\label{sec:Setup}Experimental Setup}

\begin{figure}[b]
		\centering
		\includegraphics[width=0.48\textwidth,clip]{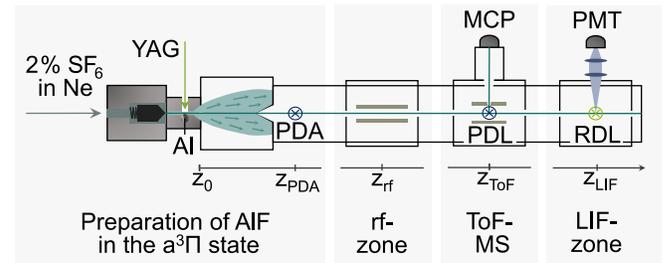}
	\caption{\label{fig:Setup} Scheme of the molecular beam setup. The arrangement of the rf interaction region, the Time-of-Flight mass spectrometer (ToF-MS) and the Laser Induced Fluorescence (LIF) zone can readily be changed.}
\end{figure}

The experimental setup is shown schematically in Figure~\ref{fig:Setup}, and has been described in detail elsewhere.\cite{nicoleastate} A gas mixture of 2~\% SF$_6$ seeded in neon with a backing pressure of 3~bar is released from a pulsed valve, operated at 10~Hz, and passes through a narrow channel. The radiation of a pulsed Nd:YAG laser (5 ns pulse duration, 16~mJ pulse energy, 1064~nm) is mildly focused on a rotating aluminum rod, injecting aluminum atoms in this channel. AlF is produced in a chemical reaction of the laser ablated aluminum atoms and SF$_6$. The molecules are cooled while expanding from the narrow channel into vacuum, reaching a translational temperature of about 3~K and a rotational temperature of about 5~K. The vibrational cooling is comparatively inefficient, and as a result about 1~\% of the population occupies the X$^1\Sigma^+, v=5$ state. This is a sufficient starting point for the experiments presented here. The molecules pass through a skimmer and then enter the differentially pumped preparation region. Here, the radiation from a TiSa-seeded, frequency doubled pulsed dye amplifier (PDA) crosses the molecular beam perpendicularly and excites the molecules on selected rotational lines of the a$^3\Pi, v=5$ $\leftarrow$ X$^1\Sigma^+, v=5$ band around 365~nm.

The lifetime of the molecules in the a$^3\Pi, v=5$ state is several ms\cite{nicoleastate} and their forward velocity is about 750~m/s. Therefore, manipulation and detection can take place in spatially separated regions. By passing through a parallel plates transmission line, the molecules can be exposed to rf radiation to populate selected $F$ levels in the a$^3\Pi, v=5$ state. The metastable molecules can be monitored by ionization or LIF detection via the b$^3\Sigma^+, v=5$ state in regions further downstream. Rotationally resolved electronic spectra are recorded via (1+1$'$)~REMPI using a pulsed dye laser around 580~nm (0.05~cm$^{-1}$ bandwidth, several $\mu$J pulse energy) for resonant excitation, followed by a counter-propagating pulsed dye laser around 280~nm (several mJ pulse energy) for ionization. The time delay between the excitation and ionization laser pulse can be varied with ns precision. Parent AlF$^+$ ions are detected in a compact, linear Time-of-Flight mass spectrometer (ToF-MS) of the Wiley-McLaren type. For LIF detection, a narrowband cw ring dye laser is used to resonantly excite the molecules to the A$^1\Pi, v=6$ // b$^3\Sigma^+, v=5$ system. The UV fluorescence to the X$^1\Sigma^+$ state, allowed for by the singlet character of the excited levels, is detected with a photomultiplier tube (PMT). The absolute frequency of the ring dye laser is recorded with a wavemeter (HighFinesse WS8) to an absolute accuracy of 10~MHz.

\section{\label{sec:Rot}Rotationally Resolved Spectroscopy}

\begin{figure}
		\centering
		\includegraphics[width=0.48\textwidth,clip]{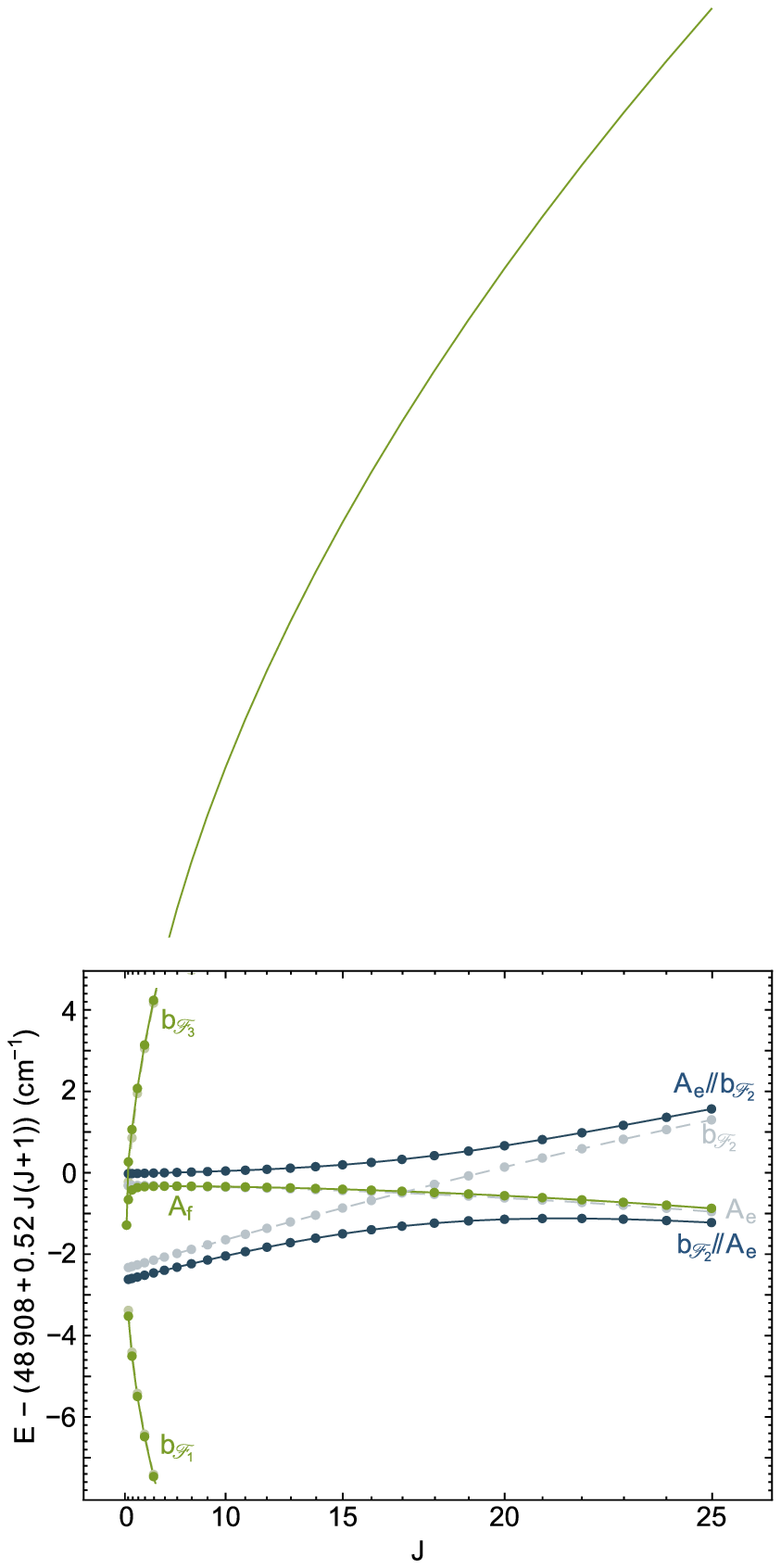}
	\caption{\label{fig:PertRot1}Overview of the A$^1\Pi, v=6$ // b$^3\Sigma^+, v=5$ perturbation. The energy of the levels of the b$^3\Sigma^+, v=5$ and A$^1\Pi, v=6$ states, reduced by $\left( 48908.0+ 0.52 \, J(J+1 )\right)$~cm$^{-1}$, are plotted as a function of $J$, on a scale linear in $J(J+1)$. The $e$ levels are plotted in blue, the $f$ levels in green. Solid (dashed) curves connect the (un)perturbed levels.
}
\end{figure}

The $v=0$ level of the A$^1\Pi$ state of AlF is located about 855~cm$^{-1}$ below the $v=0$ level of the b$^3\Sigma^+$ state.\cite{1974barrow} This spacing is only slightly larger than the vibrational spacing in the A$^1\Pi$ state, locating the $v+1$ level of the A$^1\Pi$ state close in energy to the $v$ level in the b$^3\Sigma^+$ state. As the vibrational frequency in the A$^1\Pi$ state is somewhat larger than the one in the b$^3\Sigma^+$ state, these vibrational levels come gradually closer together with increasing $v$ and eventually cross. The energy gap is the smallest for the A$^1\Pi,v=6$ and the b$^3\Sigma^+,v=5$ states, for which the lowest rotational levels are schematically depicted in the top right part of Figure~\ref{fig:EnergyScheme}.

For an overview of the interaction between the rotational levels in the A$^1\Pi,v=6$ and b$^3\Sigma^+,v=5$ states, we first neglect the hyperfine structure as well as splittings due to $\Lambda$-doubling, spin-spin and spin-rotation interaction. We assume that $J$, the quantum number belonging to the total angular momentum $\mathbf{J}$, is a good quantum number. The energy of the unperturbed rotational levels in the A$^1\Pi, v=6$ state (Hund's case (a)) is given by $B_{\text{A}6} J(J+1)$, where $B_{\text{A}6}$ is the rotational constant of the A$^1\Pi, v=6$ state. Each $J$ level consists of a positive and a negative component of the $\Lambda$-doublet whose splitting is neglected, i.e. the $e$ levels and $f$ levels are assumed to be degenerate. In the b$^3\Sigma^+, v=5$ state (Hund's case (b)), $\mathbf{J}$ is the vectorial sum of the end-over-end rotation $\mathbf{N}$ and the total spin of the electrons $\mathbf{S}$. The expression for the unperturbed rotational energy is given by $B_{\text{b}5} N(N+1)$, where $B_{\text{b}5}$ is the rotational constant of the b$^3\Sigma^+, v=5$ state. In the absence of spin-spin and spin-rotation interaction, the levels with $J=N+1$, $J=N$ and $J=N-1$, referred to as $\mathcal{F}_1$, $\mathcal{F}_2$ and $\mathcal{F}_3$ levels, respectively, are degenerate. The $\mathcal{F}_1$ and $\mathcal{F}_3$ levels are $f$ levels whereas the $\mathcal{F}_2$ levels are $e$ levels. Using the spectroscopic constants as deduced by Barrow {\it et~al.} the unperturbed $J=1$ level of the A$^1\Pi, v=6$ state is only 0.09(3) cm$^{-1}$ below the unperturbed $N=2$ level of the b$^3\Sigma^+, v=5$ state.\cite{1974barrow}

Interaction due to spin-orbit coupling occurs between levels that have the same values of $J$ and the same $e$ or $f$ character. For the $e$ levels, the energy is found as the Eigenvalues of the $2\times2$ matrix \cite{kovacs}
\begin{equation}
\left ( \begin{array}{cc}
E_6(\text{A}_{e,J})  & -\frac{\xi_{6,5}}{\sqrt{2}}  \\
 -\frac{\xi_{6,5}}{\sqrt{2}} & E_5({\text{b}_{\mathcal{F}_2,J}}) 
\end{array} \right )
\label{eq:pertmatrix_e}
\end{equation}
where $E_6(\text{A}_{e,J})$ and $E_5({\text{b}_{\mathcal{F}_2,J}})$ are the energies of the unperturbed $J,e$ levels of the A$^1\Pi, v=6$ state and b$^3\Sigma^+, v=5$ state, respectively. The parameter $\xi_{6,5}$ is the spin-orbit interaction constant for these two states. As $J=N$ for the $e$ levels in the b$^3\Sigma^+$ state, and as the rotational constants of the interacting states only differ by 1~\%, the effect of this perturbation is detectable over a wide range of $J$ levels, as shown with the blue dots in Figure~\ref{fig:PertRot1}. From this perturbation between the $e$ levels, Barrow {\it et~al.}\cite{1974barrow} determined the relative energies of the singlet and the triplet levels and found the spin-orbit interaction constant $\xi_{6,5}$ to be about 1.16(1)~cm$^{-1}$.

\begin{figure}
		\centering
		\includegraphics[width=0.48\textwidth,clip]{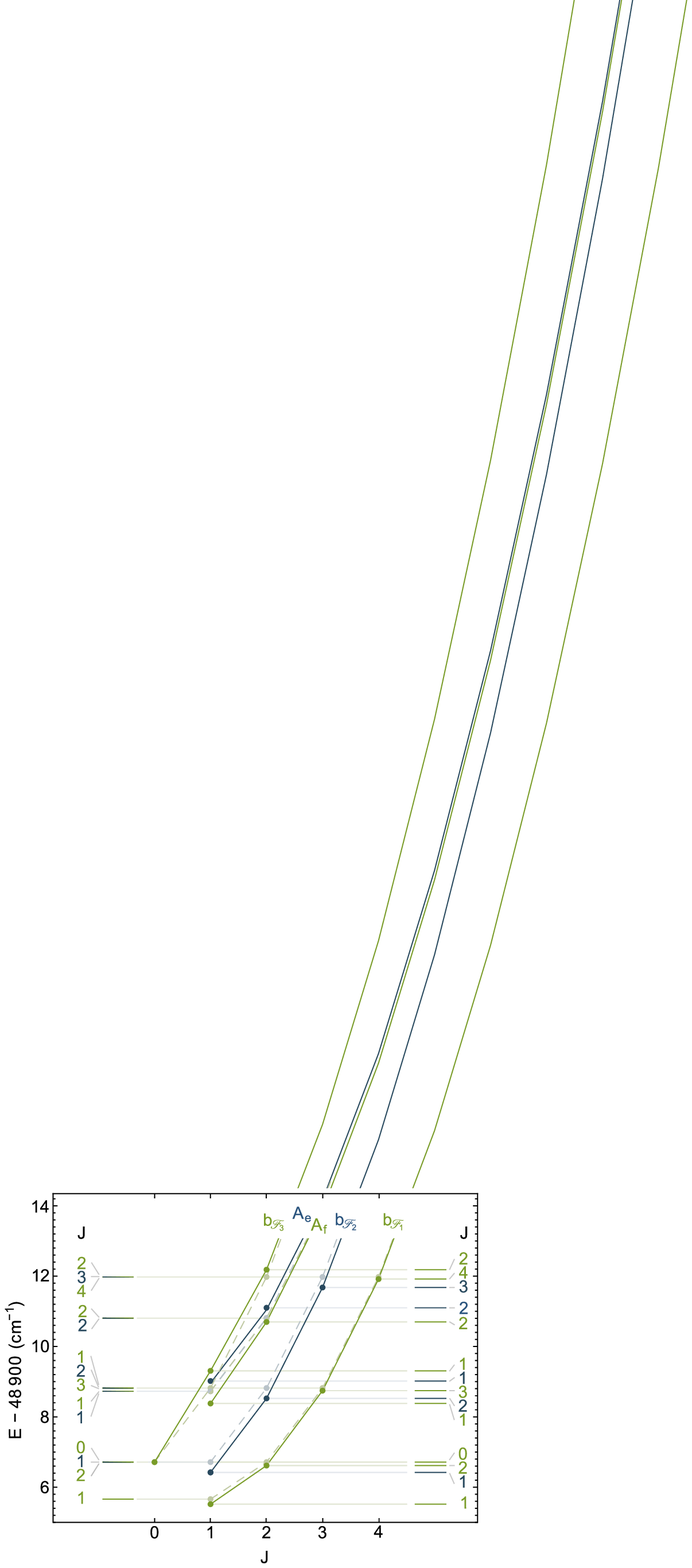}
	\caption{\label{fig:PertRot2}
	Energy of the levels of the b$^3\Sigma^+, v=5$ and A$^1\Pi, v=6$ states minus $48900$ cm$^{-1}$ as a function of $J$, for $J=0-4$. The $e$ levels are plotted in blue, the $f$ levels in green. Solid (dashed) curves connect the (un)perturbed levels.
}
\end{figure}

Both the $\mathcal{F}_1$ and $\mathcal{F}_3$ levels of the b$^3\Sigma^+$ state are $f$ levels, so the energies of the perturbed $f$ levels for a given $J$ are found as the Eigenvalues of the $3\times3$ matrix \cite{kovacs}
\begin{equation}
\left ( \begin{array}{ccc}
E_6(\text{A}_{f,J})  & 
\sqrt{\frac{J+1}{4J+2}}\xi_{6,5} &
\sqrt{\frac{J}{4J+2}}\xi_{6,5} \\
\sqrt{\frac{J+1}{4J+2}}\xi_{6,5} &
E_5({\text{b}_{\mathcal{F}_1,J}}) &
0 \\
\sqrt{\frac{J}{4J+2}}\xi_{6,5} &
0 &
E_5({\text{b}_{\mathcal{F}_3,J}})
\end{array} \right )
\label{eq:pertmatrix_f}
\end{equation}
where $E_6(\text{A}_{f,J})$, $E_5({\text{b}_{\mathcal{F}_1,J}})$ and $E_5({\text{b}_{\mathcal{F}_3,J}})$ are the energies of the unperturbed $f$ levels of the A$^1\Pi, v=6$ state and of the $\mathcal{F}_1$ and $\mathcal{F}_3$ levels of the b$^3\Sigma^+, v=5$ state, respectively. Due to the different $J$-dependence of the rotational energy of the interacting $f$ levels, their interaction is localized to the lowest values of $J$, almost like an accidental resonance, as shown by the green dots in Figure~\ref{fig:PertRot1}.

In Figure~\ref{fig:PertRot2} the effect of the perturbation on the lowest rotational levels in the A$^1\Pi, v=6$ and b$^3\Sigma^+, v=5$ states is shown in more detail. On the left side of this Figure, the calculated energies of the lowest unperturbed levels are shown, up to $N=3$ in the b$^3\Sigma^+, v=5$ state. The perturbation lifts the degeneracy of the $J$ levels and on the right side the resulting $J$ level structure is shown. The shift for the $J=1$ levels is the largest and up to 0.5~cm$^{-1}$; the magnitude of the shift decreases for higher values of $J$. The $J=0$ level of the b$^3\Sigma^+, v=5$ state is the only one that is not affected, as there is no level in the A$^1\Pi$ state to interact with.

\begin{figure*}
		\centering
		\includegraphics[width=1\textwidth]{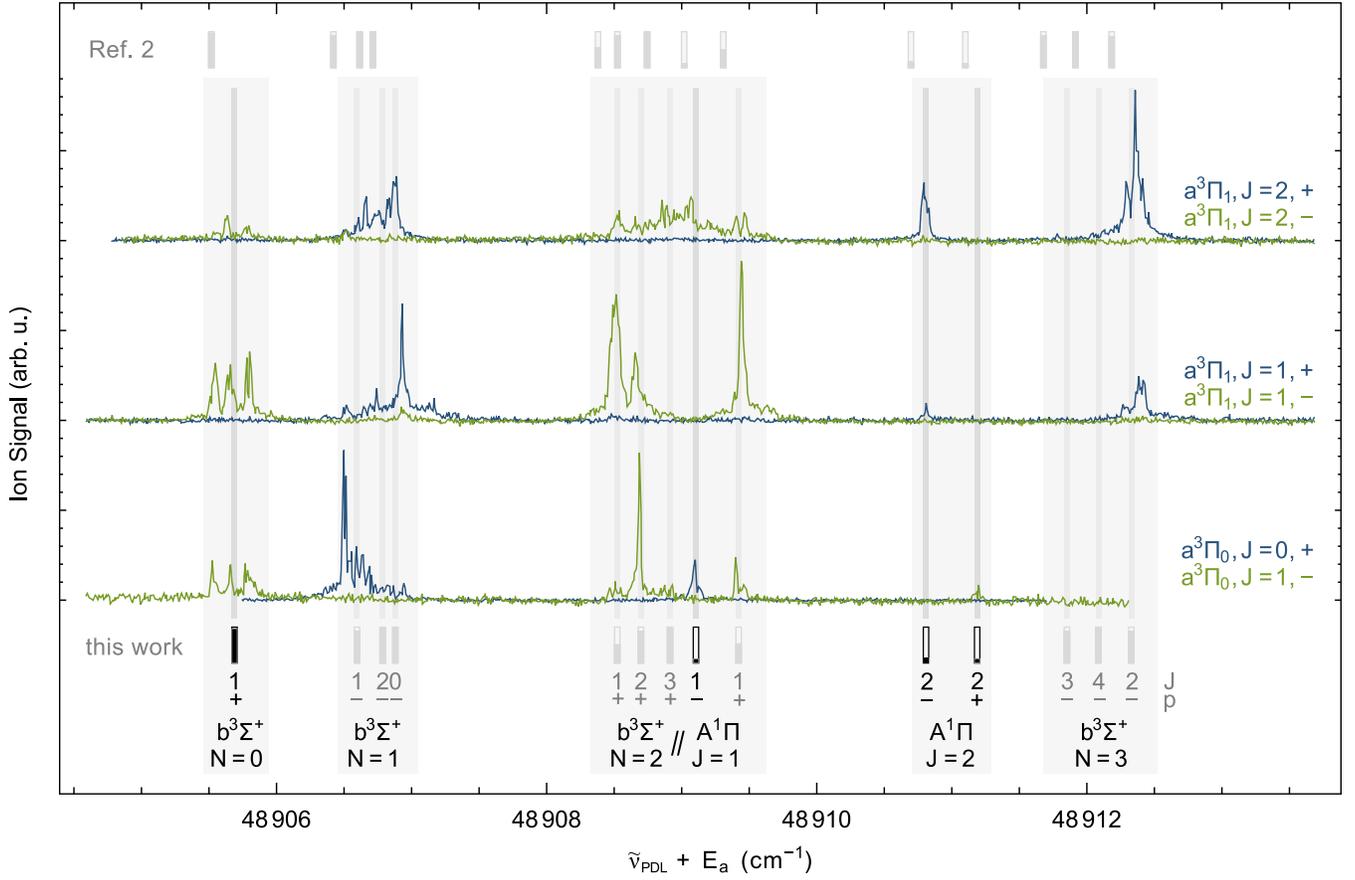}
	\caption{\label{fig:PDL}Overview excitation spectra to the lowest rotational levels of the perturbed A$^1\Pi, v=6$ // b$^3\Sigma^+, v=5$ system, recorded via (1+1$'$)~REMPI. The spectra originate from single rotational levels in the a$^3\Pi, v=5$ state at energy $E_\text{a}$ and that are indicated on the right. The spectra are plotted such that the energy level structure in the excited state, shown as grey bars near the bottom, can be directly recognized.
}
\end{figure*}

%%new (Gerard)
Overview excitation spectra from selected rotational levels in the a$^3\Pi, v=5$ state to the lowest rotational levels of the perturbed A$^1\Pi, v=6$ // b$^3\Sigma^+, v=5$ system are shown in Figure~\ref{fig:PDL}. These spectra are recorded via (1+1$'$)~REMPI, using a time delay between the excitation and ionization laser of less than 10~ns, and the spectral resolution is limited by the bandwidth of the laser. Using the known energies of the rotational levels in the a$^3\Pi, v=5$ state, the spectra are plotted such that the horizontal axis represents the absolute energies of the perturbed levels. The fourteen energy levels given on the right in Figure~\ref{fig:PertRot2} are indicated as grey bars above the spectra in Figure~\ref{fig:PDL}. The filled fraction of the grey bars corresponds to the fraction of triplet character of the corresponding $J$ levels.

\begin{longtable}{dddccccccccc}

\caption{\label{tab:linelist}Observed A$^1\Pi, v=6$ // b$^3\Sigma^+, v=5$ $\leftarrow$ a$^3\Pi, v=5$ transition frequencies, difference between the observed and calculated frequencies (all given in cm$^{-1}$), spectroscopic labels of the levels in the upper (primed) and lower (unprimed) state and the fraction of triplet ($f_{\text{b}5}$) character of the upper level.}
\\
\hline
\\[-0.3cm]
 \multicolumn{1}{c}{$\tilde{\nu}_{\text{exp}}$} &
\multicolumn{1}{c}{$\tilde{\nu}_{\text{exp}}-\tilde{\nu}_{\text{calc}}$} &
      \multicolumn{1}{c}{$f_{\text{b}5}$}&
           \multicolumn{1}{c}{$J'$}& 
             \multicolumn{1}{c}{$F'$}&
            \multicolumn{1}{c} {$p'$}&
            \multicolumn{1}{c} {$n'$}&
            \multicolumn{1}{c} {$\Omega$}&
            \multicolumn{1}{c}  {$J$}& 
            \multicolumn{1}{c} {$F$}&
            \multicolumn{1}{c} {$p$}&
             \multicolumn{1}{c}{$n$}\\[0.1cm]
\endfirsthead
\multicolumn{12}{c}%
        {{Table \thetable\ Continued from previous column.}} \\
\hline\\[-0.3cm]
 \multicolumn{1}{c}{$\tilde{\nu}_{\text{exp}}$} &
\multicolumn{1}{c}{$\tilde{\nu}_{\text{exp}}-\tilde{\nu}_{\text{calc}}$} &
      \multicolumn{1}{c}{$f_{\text{b}5}$}&
          \multicolumn{1}{c}{$J'$}& 
             \multicolumn{1}{c}{$F'$}&
            \multicolumn{1}{c} {$p'$}&
            \multicolumn{1}{c} {$n'$}&
            \multicolumn{1}{c} {$\Omega$}&
            \multicolumn{1}{c}  {$J$}& 
            \multicolumn{1}{c} {$F$}&
            \multicolumn{1}{c} {$p$}&
             \multicolumn{1}{c}{$n$}\\[0.1cm]
\hline
\endhead
\hline

17236.4122 & 0.0002 & 0.961 & 1 & 2 & $+$ & 1 & 1 & 1 & 3 & $-$ & 1 \\
17236.4198 & -0.0001 & 0.961 & 1 & 2 & $+$ & 1 & 1 & 1 & 2 & $-$ & 1 \\
17236.4369 & 0.0000 & 0.960 & 1 & 1 & $+$ & 1 & 1 & 1 & 2 & $-$ & 1 \\
17236.4416 & -0.0002 & 0.960 & 1 & 1 & $+$ & 1 & 1 & 1 & 1 & $-$ & 7 \\
17236.5228 & -0.0000 & 0.958 & 1 & 2 & $+$ & 2 & 1 & 1 & 3 & $-$ & 1 \\
17236.5232 & -0.0001 & 0.958 & 1 & 3 & $+$ & 1 & 1 & 1 & 4 & $-$ & 1 \\
17236.5294 & 0.0004 & 0.958 & 1 & 2 & $+$ & 2 & 1 & 1 & 2 & $-$ & 1 \\
17236.5339 & -0.0001 & 0.958 & 1 & 2 & $+$ & 2 & 1 & 1 & 1 & $-$ & 7 \\
17236.5407 & 0.0001 & 0.958 & 1 & 3 & $+$ & 1 & 1 & 1 & 2 & $-$ & 1 \\
17236.6592 & -0.0002 & 0.954 & 1 & 3 & $+$ & 2 & 1 & 1 & 3 & $-$ & 1 \\
17236.6656 & -0.0000 & 0.954 & 1 & 3 & $+$ & 2 & 1 & 1 & 2 & $-$ & 1 \\
17236.6851 & 0.0002 & 0.954 & 1 & 4 & $+$ & 1 & 1 & 1 & 4 & $-$ & 1 \\
17236.6946 & 0.0002 & 0.954 & 1 & 4 & $+$ & 1 & 1 & 1 & 3 & $-$ & 1 \\
17237.4064 & 0.0014 & 0.774 & 2 & 2 & $+$ & 5 & 1 & 2 & 3 & $-$ & 1 \\
17237.4097 & 0.0004 & 0.774 & 2 & 2 & $+$ & 5 & 1 & 2 & 1 & $-$ & 8 \\
17237.4192 & 0.0010 & 0.779 & 2 & 3 & $+$ & 5 & 1 & 2 & 4 & $-$ & 1 \\
17237.4257 & 0.0010 & 0.779 & 2 & 3 & $+$ & 5 & 1 & 2 & 2 & $-$ & 1 \\
17237.4323 & 0.0013 & 0.820 & 2 & 4 & $+$ & 3 & 1 & 2 & 5 & $-$ & 1 \\
17237.4323 & -0.0001 & 0.833 & 2 & 3 & $+$ & 6 & 1 & 2 & 4 & $-$ & 1 \\
17237.4400 & -0.0001 & 0.820 & 2 & 4 & $+$ & 3 & 1 & 2 & 3 & $-$ & 1 \\
17237.4400 & 0.0012 & 0.833 & 2 & 3 & $+$ & 6 & 1 & 2 & 2 & $-$ & 1 \\
17237.4531 & -0.0006 & 0.892 & 2 & 4 & $+$ & 4 & 1 & 2 & 3 & $-$ & 1 \\
17237.4579 & 0.0006 & 0.894 & 2 & 5 & $+$ & 1 & 1 & 2 & 4 & $-$ & 1 \\
17237.5197 & -0.0011 & 0.986 & 3 & 0 & $+$ & 2 & 1 & 2 & 1 & $-$ & 8 \\
17237.5635 & -0.0009 & 0.979 & 3 & 1 & $+$ & 6 & 1 & 2 & 1 & $-$ & 9 \\
17237.5782 & -0.0012 & 0.969 & 3 & 2 & $+$ & 7 & 1 & 2 & 2 & $-$ & 1 \\
17237.6067 & 0.0001 & 0.971 & 3 & 2 & $+$ & 8 & 1 & 2 & 1 & $-$ & 9 \\
17237.6263 & 0.0001 & 0.962 & 3 & 3 & $+$ & 7 & 1 & 2 & 3 & $-$ & 1 \\
17237.6566 & -0.0004 & 0.970 & 3 & 3 & $+$ & 8 & 1 & 2 & 2 & $-$ & 1 \\
17237.7118 & 0.0002 & 0.976 & 3 & 4 & $+$ & 6 & 1 & 2 & 3 & $-$ & 1 \\
17237.7371 & -0.0001 & 0.974 & 3 & 5 & $+$ & 2 & 1 & 2 & 4 & $-$ & 1 \\
17237.7663 & 0.0000 & 0.985 & 3 & 5 & $+$ & 3 & 1 & 2 & 4 & $-$ & 1 \\
17237.7899 & -0.0001 & 0.985 & 3 & 6 & $+$ & 1 & 1 & 2 & 5 & $-$ & 1 \\
17239.3590 & 0.0003 & 0.620 & 1 & 4 & $+$ & 2 & 1 & 1 & 4 & $-$ & 1 \\
17239.4221 & -0.0003 & 0.576 & 1 & 2 & $+$ & 4 & 1 & 1 & 3 & $-$ & 1 \\
17239.4367 & -0.0004 & 0.526 & 1 & 1 & $+$ & 2 & 1 & 1 & 1 & $-$ & 7 \\
17240.2834 & 0.0002 & 0.515 & 1 & 4 & $+$ & 7 & 1 & 1 & 4 & $-$ & 1 \\
17240.2864 & -0.0003 & 0.519 & 1 & 3 & $+$ & 9 & 1 & 1 & 3 & $-$ & 1 \\
17240.2926 & -0.0001 & 0.515 & 1 & 4 & $+$ & 7 & 1 & 1 & 3 & $-$ & 1 \\
17240.2926 & -0.0003 & 0.519 & 1 & 3 & $+$ & 9 & 1 & 1 & 2 & $-$ & 1 \\
17240.3177 & 0.0004 & 0.561 & 1 & 3 & $+$ & 10 & 1 & 1 & 4 & $-$ & 1 \\
17240.3177 & -0.0006 & 0.559 & 1 & 2 & $+$ & 9 & 1 & 1 & 3 & $-$ & 1 \\
17240.3251 & 0.0006 & 0.559 & 1 & 2 & $+$ & 9 & 1 & 1 & 2 & $-$ & 1 \\
17240.3293 & -0.0002 & 0.559 & 1 & 2 & $+$ & 9 & 1 & 1 & 1 & $-$ & 7 \\
17240.3345 & -0.0001 & 0.561 & 1 & 3 & $+$ & 10 & 1 & 1 & 2 & $-$ & 1 \\
17240.3466 & 0.0005 & 0.584 & 1 & 1 & $+$ & 7 & 1 & 1 & 2 & $-$ & 1 \\
17240.3515 & 0.0002 & 0.592 & 1 & 2 & $+$ & 10 & 1 & 1 & 3 & $-$ & 1 \\
17240.3515 & 0.0004 & 0.584 & 1 & 1 & $+$ & 7 & 1 & 1 & 1 & $-$ & 7 \\
17240.3594 & 0.0002 & 0.592 & 1 & 2 & $+$ & 10 & 1 & 1 & 2 & $-$ & 1 \\
17286.1879 & 0.0001 & 0.904 & 1 & 2 & $-$ & 1 & 0 & 0 & 3 & $+$ & 1 \\
%17286.1879 & 0.0001 & 0.904 & 1 & 2 & $-$ & 1 & 0 & 0 & 3 & $+$ & 1 \\
%17286.2099 & -0.0000 & 0.910 & 1 & 3 & $-$ & 1 & 0 & 0 & 3 & $+$ & 1 \\
17286.2099 & -0.0000 & 0.910 & 1 & 3 & $-$ & 1 & 0 & 0 & 3 & $+$ & 1 \\
17286.2123 & -0.0002 & 0.899 & 1 & 1 & $-$ & 1 & 0 & 0 & 2 & $+$ & 1 \\
17286.2271 & 0.0001 & 0.905 & 1 & 2 & $-$ & 2 & 0 & 0 & 2 & $+$ & 1 \\
%17286.2274 & 0.0004 & 0.905 & 1 & 2 & $-$ & 2 & 0 & 0 & 2 & $+$ & 1 \\
17286.3087 & -0.0003 & 0.896 & 1 & 3 & $-$ & 2 & 0 & 0 & 2 & $+$ & 1 \\
%17286.3087 & -0.0002 & 0.896 & 1 & 3 & $-$ & 2 & 0 & 0 & 2 & $+$ & 1 \\
%17286.3160 & -0.0004 & 0.899 & 1 & 4 & $-$ & 1 & 0 & 0 & 3 & $+$ & 1 \\
17286.3166 & 0.0001 & 0.899 & 1 & 4 & $-$ & 1 & 0 & 0 & 3 & $+$ & 1 \\
17288.7906 & 0.0002 & 0.106 & 1 & 2 & $-$ & 6 & 0 & 0 & 3 & $+$ & 1 \\
17288.7945 & -0.0000 & 0.107 & 1 & 1 & $-$ & 4 & 0 & 0 & 2 & $+$ & 1 \\
17288.8150 & -0.0001 & 0.115 & 1 & 3 & $-$ & 7 & 0 & 0 & 2 & $+$ & 1 \\
17288.8197 & 0.0000 & 0.117 & 1 & 4 & $-$ & 4 & 0 & 0 & 3 & $+$ & 1 \\
\hline
\end{longtable}

Although the overall pattern of the calculated energy levels,
shown above the spectra,
agrees with the observations, a detailed inspection shows that the absolute frequency and splitting of the levels is slightly off. It appears that both the absolute energies of the unperturbed levels of the A$^1\Pi, v=6$ and b$^3\Sigma^+, v=5$ states and the spin-orbit interaction parameter $\xi_{6,5}$ need to be slightly adjusted. 
The energy levels calculated with the improved parameters are shown underneath the spectra in Figure~\ref{fig:PDL}.
To find these improved parameters, 
the three transitions to the perturbed levels with predominant singlet character (indicated in black) are used.
As the hyperfine structure of these transitions is within the bandwidth of the laser, their center frequencies can be accurately determined. 
The hyperfine structure of the perturbed b$^3\Sigma^+, v=5,N=0,J=1,+$ level results in three groups of hyperfine levels, spanning over 0.3~cm$^{-1}$, and will be similar to that of the corresponding level in the b$^3\Sigma^+, v=0$ state.\cite{max} From this we know that the center of gravity of this level, i.e.\ the frequency of this level in the absence of fine and hyperfine structure, is about 0.04~cm$^{-1}$ above the central group of hyperfine levels (also indicated in black). By defining the center frequencies of the transitions to the predominantly singlet and triplet levels in this way, a better agreement with the observations is obtained when the  
separation between the unperturbed A$^1\Pi, v=6$ and b$^3\Sigma^+, v=5$ states is about 0.07~cm$^{-1}$ larger than given by Barrow {\it et~al.}\cite{1974barrow}, whereas the spin-orbit interaction parameter is $\xi_{6,5}=1.12$~cm$^{-1}$, slightly smaller than given by Barrow {\it et~al.}\cite{1974barrow}
In addition, the absolute energies of both the  A$^1\Pi, v=6$ and b$^3\Sigma^+, v=5$ states needed to be shifted somewhat upwards.

The intensity of the lines in the spectra correlates with the amount of triplet character of the levels for two reasons: First, the transition intensity is determined by the amount of triplet character. Second, the radiative lifetime of the levels in the upper state is longer, i.e. these are detected more efficiently in the (1+1$'$)~REMPI process, when the amount of triplet character is larger. Singlet-triplet mixing is seen to be particularly pronounced for the positive parity, $J=1$ levels belonging to the A$^1\Pi, v=6, J=1$ // b$^3\Sigma^+, v=5, N=2$ system. The parity of the levels is well defined. The apparent violation of the parity selection rule in some of the spectra is attributed to mixing of closely spaced, opposite parity hyperfine levels in the a$^3\Pi_1, v=5$ state due to stray electric fields in the ionization detection region. The modeling of the perturbation as given here is seen to be sufficiently accurate for understanding the observed transitions; additional splittings and energy level shifts are due to hyperfine, spin-spin and spin-rotation interactions, as discussed in the next Section.

%%%%%%%%%%%%%%%%%%%%%%%%%%%%%%%%%%%%%%%%%%%%%%%%%%%%%%%%%%%%%%%%%%%%%%%%%%%%%%%%%%%%%%%%%%%%%%%%%%%%%%%%%%%%%%%%%%%%%%%%%%%%%%%%%%%%%%%%%%%%%%%%%%%%%%%%%%%%%%%%%%%%%%%%%%%%%%%%%%%%%%%
\section{\label{sec:Hyperfine}Hyperfine Resolved Spectroscopy}
\begin{figure*}
		\centering
		\includegraphics[width=1\textwidth]{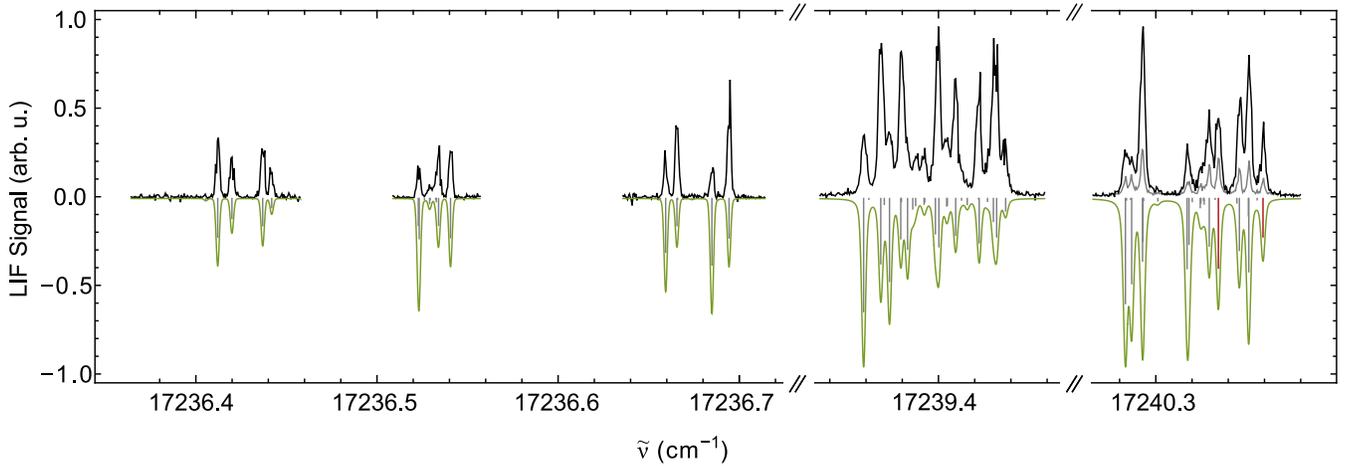}
	\caption{\label{fig:hyperfine1}Hyperfine resolved LIF spectrum from the a$^3\Pi_1, v=5, J=1,-$ level to the various positive parity hyperfine components of the perturbed  A$^1\Pi, v=6$ // b$^3\Sigma^+, v=5$ system.
	Measured spectra pointing upwards in black, simulated spectrum pointing downwards in green.
	The spectroscopic labeling of the different transitions is given in Table~\ref{tab:linelist}.
}
\end{figure*}

The fine structure and hyperfine structure in the $v=0$ levels of the A$^1\Pi$ and b$^3\Sigma^+$ states has recently been precisely measured and analyzed.\cite{alf,max} The model presented in those studies is used to describe the structure in the excited vibrational levels as well, augmented with the perturbation between the electronic states as described in the previous Section. Whereas the span of the hyperfine structure in the A$^1\Pi, v=0$ state is less than 0.02~cm$^{-1}$, it is about 0.30 cm$^{-1}$ in the b$^3\Sigma^+, v=0$ state. The latter is much larger than the splitting due to spin-rotation interaction, resulting in $J$ not being a good quantum number; only for the single $N=0$ level in the b$^3\Sigma^+, v=0$ state, $J=1$ is well defined. When the hyperfine interaction is strong compared to the spin-rotation interaction, it is useful to introduce an intermediate angular momentum $\mathbf{G}$ = $\mathbf{I_{Al}}$ + $\mathbf{S}$, where $\mathbf{I_{Al}}$ is the spin of the aluminum nucleus. This angular momentum $\mathbf{G}$ then couples to $\mathbf{N}$ and $\mathbf{I_{F}}$, the spin of the fluorine nucleus, to the total angular momentum $\mathbf{F}$. This approach has been followed and detailed in the study on the b$^3\Sigma^+, v=0$ state.\cite{max}

In the $v=5$ level of the b$^3\Sigma^+$ state, the perturbation with the A$^1\Pi$ state results in a splitting of the $J$ levels belonging to the same $N$ that can be considerably larger than the hyperfine splitting. Interestingly, this makes $J$ a rather good quantum number for all the levels belonging to $N=0-3$, explaining the overall good agreement between the measured and calculated energies of the $J$ levels as shown in Figure~\ref{fig:PDL}. 

The resolution in the spectra shown in Figure~\ref{fig:PDL} is sufficient to partly resolve the fine  and hyperfine structure, as is best seen in the transitions to the $J=1$, $\mathcal{F}_1$ level of $N=0$ in the b$^3\Sigma^+, v=5$ state, for which three (groups of) lines can be recognized. Higher resolution spectra are recorded using a narrowband cw ring dye laser (< 1~MHz bandwidth, $1 - 20$~mW) for excitation on selected transitions of the A$^1\Pi, v=6$ // b$^3\Sigma^+, v=5$ $\leftarrow$ a$^3\Pi, v=5$ band, followed by far off-resonant detection of the laser induced fluorescence back to the X$^1\Sigma^+$ state. On the low frequency side of the spectrum shown in Figure~\ref{fig:hyperfine1}, the hyperfine resolved transitions from the a$^3\Pi_1,v=5, J=1$ level to the b$^3\Sigma^+, v=5, N=0$ level are shown. The width of the spectral lines is about 70~MHz, mainly determined by residual Doppler broadening in the molecular beam. The hyperfine structure in the a$^3\Pi_1,v=5, J=1$ level is known to about 20~kHz accuracy from earlier measurements.\cite{nicoleastate} The perturbation with the A$^1\Pi, v=6$ state results mainly in an overall shift to lower energy of the b$^3\Sigma^+, v=5, N=0$ hyperfine levels. This high resolution spectrum is therefore well suited to constrain the fine structure and hyperfine structure parameters in the b$^3\Sigma^+, v=5$ state. 

\begin{figure}
		\centering
		\includegraphics[width=0.48\textwidth,clip]{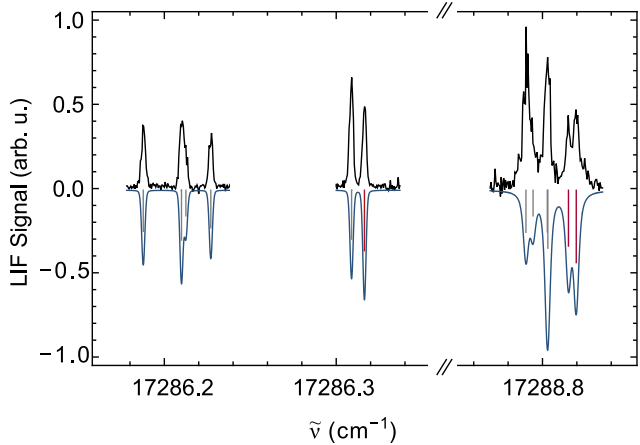}
	\caption{\label{fig:hyperfine2}
	Hyperfine resolved LIF spectrum from the a$^3\Pi_0, v=5, J=0,+$ level to the various negative parity hyperfine components of the perturbed A$^1\Pi, v=6$ // b$^3\Sigma^+, v=5$ system.
	Measured spectra pointing upwards in black, simulated spectrum pointing downwards in blue. 
	The spectroscopic labeling of the different transitions is given in Table~\ref{tab:linelist}.
}
\end{figure}

The spectra shown in Figure~\ref{fig:hyperfine2} are recorded from the positive parity, $J=0$ level in the a$^3\Pi_0$ state. This $J=0$ level has two hyperfine components with a separation of only a few MHz (3~MHz for $v=0$, expected to be similar for $v=5$)\cite{alf} such that the spectra basically directly reflect the level structure in the excited state. At low frequency, the transition to the $N=1$ level of the b$^3\Sigma^+, v=5$ state is shown, and about 2.5 cm$^{-1}$ higher the transition to the negative parity, $J=1$ level of the A$^1\Pi, v=6$ state is shown. The latter level has predominantly singlet character, and the spectral lines are broadened due to its short lifetime. The overall hyperfine structure of this $J=1$ level is similar to what has been observed for the $J=1$ level in the A$^1\Pi, v=0$ state, but it is expanded on the energy scale due to the perturbation by about a factor two. The observed hyperfine splitting of this level is therefore well suited to constrain the hyperfine structure parameters in the A$^1\Pi, v=6$ state. 

The two spectra shown on the high frequency side of Figure~\ref{fig:hyperfine1} are recorded from the a$^3\Pi_1,v=5, J=1$ level, reaching the strongly mixed $N=2,J=1$, positive parity levels of the A$^1\Pi, v=6$ // b$^3\Sigma^+, v=5$ system. Two different sets of measurements are shown: the black spectrum is recorded with the same residual Doppler width as the other spectra in this Figure, whereas for the recording of the grey spectrum a 4~mm wide vertical slit is inserted in the molecular beam in front of the LIF region, reducing the Doppler width by about a factor three. This spectrum is particularly sensitive to the spin-orbit interaction parameter $\xi_{6,5}$.

From the high resolution LIF spectra, a total of 59 resolved hyperfine transitions are identified. The observed spectral lines are fitted with a Voigt profile, and the center frequencies $\tilde{\nu}_{\text{exp}}$ are determined to an accuracy of better than 10~MHz. The observed frequencies  and assignments are given in Table~\ref{tab:linelist} together with the differences from the calculated values $\tilde{\nu}_{\text{calc}}$. 
The $F$ quantum number, the parity and a counting integer $n$ (starting from $n=1$ for the lowest energy level with a given $F$ and a given parity) suffice to uniquely identify each level, but the quantum number $J$ is given as well.
Primed labels are for levels in the A$^1\Pi, v=6$ // b$^3\Sigma^+, v=5$ perturbed system, unprimed ones are for levels in the a$^3\Pi_{\Omega}, v=5$ state. The fraction of triplet character of the upper levels is given as $f_{\text{b}5}$. The standard deviation of the fit is 15~MHz.

\begin{table}
\caption{Spectroscopic parameters obtained from the fit for the A$^1\Pi$ state (upper part) and for the b$^3\Sigma^+$ state (lower part). All values are in MHz. The parameters determined in this study for the $v=6$ and $v=5$ levels of these states are compared to those determined earlier for the $v=0$ levels.\cite{alf,max} The spin-orbit interaction parameter $\xi_{6,5}$ is 1.1220 cm$^{-1}$ (SD$\,\sqrt{Q}=0.0012$~cm$^{-1}$).}
\label{tab:parameters}
\begin{ruledtabular}
\begin{tabular}{ldcdc}
%\begin{tabular}{ld{2.4}d{2.0}d{2.4}d{2.0}}
&\multicolumn{2}{c}{A$^1\Pi,v=0$}&\multicolumn{2}{c}{A$^1\Pi,v=6$}\\
\cline{2-3}\cline{4-5}\\[-0.3cm]
	Parameter         & \multicolumn{1}{c}{Value\footnote{\citet{alf} (2019)}}  & \multicolumn{1}{c}{$\textrm{SD}\sqrt{\textrm{Q}}$}  & \multicolumn{1}{c}{Value}   & \multicolumn{1}{c}{$\textrm{SD}\sqrt{\textrm{Q}}$} \\
\hline\\[-0.3cm]
$B_{\text{v}}$ & 16601.9 & 0.3 & 15583.2120 & fixed \\
$D$         & 0.0318 & fixed & 0.0318 & fixed \\
$q$ & -2.94 & 0.06 & -2.94 & fixed \\
$a$(Al) & 113 & 5 & 105.3 & 5.1 \\
$eq_0Q$(Al) & 0.00 & fixed &  0.00 & fixed\\
$eq_2Q$(Al) & 40.00 & fixed &  40.00 & fixed \\
$a$(F) & 181 & 5 & 152 & 26 \\[0.18cm]
\hline
 \hline
 \\[-0.3cm]
&\multicolumn{2}{c}{b$^3\Sigma^+,v=0$}&\multicolumn{2}{c}{b$^3\Sigma^+,v=5$}\\ 
\cline{2-3}\cline{4-5}\\[-0.3cm]
	Parameter         & \multicolumn{1}{c}{Value\footnote{\citet{max} (2021)}}  & \multicolumn{1}{c}{$\textrm{SD}\sqrt{\textrm{Q}}$}  & \multicolumn{1}{c}{Value}   & \multicolumn{1}{c}{$\textrm{SD}\sqrt{\textrm{Q}}$} \\
\hline\\[-0.3cm]	
$B_{\text{v}}$ & 16772 & 5 & 15782.8737 & fixed \\
$\lambda$ & -919 & 18 & -652 & 38 \\
$\gamma$ & -9 & 13 & -2.5 & 4.3 \\
$b_{F}$(Al) & 1311 & 3 & 1340.8 & 1.6 \\
$c$(Al) & 73 & 18 & 86 & 9 \\
$eq_0Q$(Al) & -62 & 99 &  -49.6 & fixed\\
$b_{F}$(F) & 870 & 11 & 834.6 & 6.9 \\
$c$(F) & 305 & 53 & 401 & 35 \\
\end{tabular}
\end{ruledtabular}
\end{table}

In all cases, the simulated spectra plotted upside down underneath the experimental spectra in Figure~\ref{fig:hyperfine1} and \ref{fig:hyperfine2} reproduce the observed line positions very well. The relative line intensities are seen to differ. In the simulation it is assumed that all the hyperfine levels in the a$^3\Pi, v=5$ state are equally populated. However, this is not the case for the spectra recorded from the a$^3\Pi_1,v=5, J=1$ level, as the  bandwidth of the a$^3\Pi$ $\leftarrow$ X$^1\Sigma^+$ preparation laser is narrower than the span of the hyperfine structure in this level. The relative intensities of the lines in the recorded spectra are also influenced by the strongly varying ratio of the homogeneous (Lorentzian) to the inhomogeneous (Gaussian) contribution to the linewidth for the various spectral lines. Considering the near-saturation conditions that these measurements have been made under, this leads to varying widths of the (off-axis) velocity group that can contribute to the LIF signal, and thereby to varying signal intensities.

The spectroscopic parameters of the A$^1\Pi, v=6$ and the b$^3\Sigma^+, v=5$ states obtained from the fit, are given in Table II and are compared to the corresponding values for the $v=0$ levels determined earlier.\cite{alf,max} The value found for the spin-orbit interaction constant $\xi_{6,5}$ = 1.1220(12)~cm$^{-1}$, is slightly smaller than the value of $\sqrt2 H_0$ $\equiv$ $\xi_{6,5}$ = 1.16(1) cm$^{-1}$ found by Barrow {\it et~al.}\cite{1974barrow} Using these values of the spectroscopic parameters, the unperturbed $J=1$ level of the A$^1\Pi, v=6$ state is seen to be 0.178(4)~cm$^{-1}$ below the $N=2$ level of the b$^3\Sigma^+, v=5$ state. For the a$^3\Pi, v=5$ state, the fine structure and hyperfine structure parameters as given  elsewhere are taken.\cite{nicoleastate}
Just as done there, the spin-orbit constant $A_5$ and the spin-spin interaction constant $\lambda_5$ are kept fixed. Given the limited number of rotational levels of the a$^3\Pi, v=5$ state that are included in the fit, the parameters $A_5$, $\lambda_5$ and the $\Lambda$-doubling parameter $o_5$ are highly correlated. Only the parameter $\tilde{A}_5 = A_5 - 2\lambda_5 +(o_5+p_5+q_5)$ can be determined and by varying $o_5$ in the fit, this is found as $\tilde{A}_5$ = 47.730(1)~cm$^{-1}$. The rotational constant in the a$^3\Pi, v=5$ state is also fitted and found as $B_5=15954.1$(9.7)~MHz. Both values are in good agreement with the values of $\tilde{A}_5$ = 47.733(3)~cm$^{-1}$ and $B_5=15942.1$(6.3)~MHz found independently earlier.\cite{nicoleastate}

\section{\label{sec:Lifetimes}Lifetime Measurements}

\begin{figure}[b]
		\centering
		\includegraphics[width=0.48\textwidth,clip]{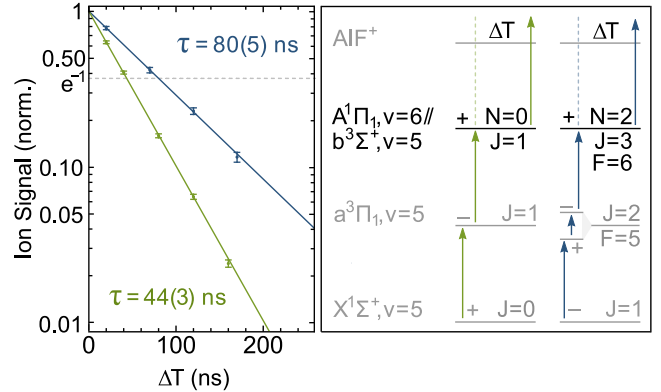}
	\caption{\label{fig:expdecay} Measurement of the AlF$^+$ ion signal as a function of the time delay $\Delta T$ between the excitation and the ionization laser pulses. The excitation schemes and the levels for which the lifetimes are measured are indicated on the right.}
\end{figure}

The fit of the hyperfine resolved transitions yields the fraction of singlet and triplet character for each of the hyperfine levels belonging to the perturbed A$^1\Pi, v=6$ // b$^3\Sigma^+, v=5$ system. To verify the fraction of triplet character, $f_{\text{b5}}(F,n,\pm)$, of the hyperfine levels, lifetime measurements are performed. The lifetime $\tau$ of a certain hyperfine level characterized by the quantum number $F$, the counting integer $n$ and the parity, is given by
\begin{equation}
\label{eq:tau}
\frac{1}{\tau(F,n,\pm)} = 
\frac{1-f_{\text{b}5}(F,n,\pm)} {\tau_{\text{A}6}}+
\frac{f_{\text{b}5}(F,n,\pm)} {\tau_{\text{b}5}}
\end{equation}
and largely differs for the different hyperfine levels. 

% A lifetime
The lifetime of the A$^1\Pi, v=0$ state has been measured to be 
$\tau_{\text{A}0}=1.90$(3)~ns.\cite{alf}
The amount of singlet-triplet mixing in this $v=0$ state is very small and its lifetime is the radiative lifetimes of the unperturbed vibrational ground state.
The found value agrees very well with the theoretically predicted value of 1.89~ns\cite{langhoff}.
These calculations predict that the lifetime of the bare A$^1\Pi,v$ states increases with $v$, and that $\tau_{\text{A}6}=2.05$~ns.

\begin{figure}
		\centering
		\includegraphics[width=0.48\textwidth,clip]{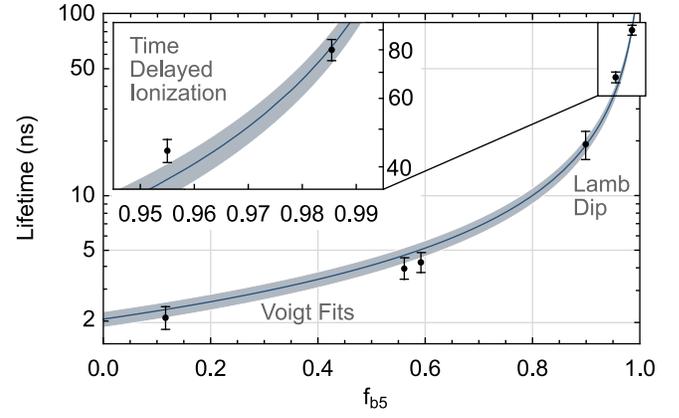}
	\caption{\label{fig:LifetimeSummary}
Measured lifetimes for six levels with a different amount of b$^3\Sigma^+$ character ($f_{\text{b}5}$).
The solid curve represents the predicted lifetime for $\tau_{\text{A}6}=2.05$~ns and $\tau_{\text{b}5}=190$~ns;
varying both values by up to 10~\% gives the shaded area. 
}
\end{figure}

% b lifetime
The lifetime of the b$^3\Sigma^+, v=0$ state is found to be $\tau_{\text{b}0}=190$(2)~ns.\cite{max} The lifetimes of the $v=1$ and $2$ states are measured in the context of this study using time-delayed ionization to be 
$\tau_{\text{b}1}=189$(4)~ns
and
$\tau_{\text{b}2}=191$(8)~ns.
For these three states, the effect on the lifetime due to the perturbation with the A$^1\Pi, v+1$ state can be neglected.
In the absence of the perturbation with the A$^1\Pi,v=6$ state, the b$^3\Sigma^+,v=5$ state can therefore be assumed to have a lifetime of $\tau_{\text{b}5}=190$~ns. 
Using these values for $\tau_{\text{A}6}$ and $\tau_{\text{b}5}$, the lifetime as a function of $f_{\text{b}5}$ is shown as the solid curve in Figure~\ref{fig:LifetimeSummary}. The shaded area corresponds to an error bar of 10~\% in both of these lifetime values. 

\begin{figure}
		\centering
		\includegraphics[width=0.48\textwidth,clip]{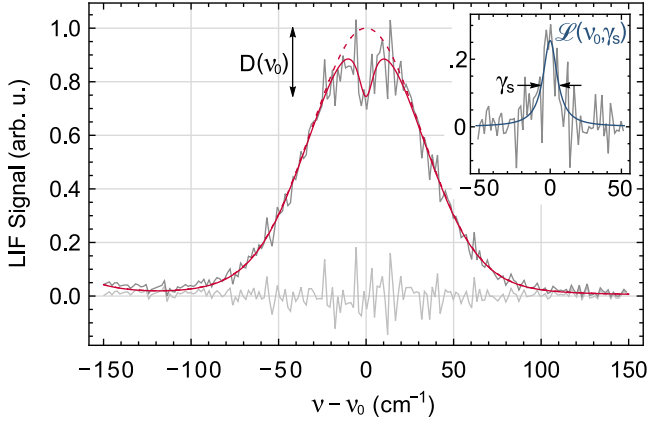}
	\caption{\label{fig:Lamb-dip}LIF intensity as a function of the laser frequency measured in a Lamb dip setup on the
 A$^1\Pi, v=6$ // b$^3\Sigma^+, v=5, N=1, J=1, F=4,-$
	$\leftarrow$
	a$^3\Pi_0, v=5, F=3, J=0,+$
	transition at  $\nu_0=17286.316$~cm$^{-1}$.
}
\end{figure}

%Voigt fits. 
The transitions shown on the high frequency side in Figure~\ref{fig:hyperfine2} reach levels that only have around 10~\% triplet character. 
The lifetime of these levels is only slightly larger than $\tau_{\text{A}6}$ and the lifetime broadening of the spectral lines is in the same range as the Doppler broadening.
Voigt fits to the observed lineshapes of the
A$^1\Pi,v=6,J=1,F=3,-$ $\leftarrow$ a$^3\Pi_0,v=5,J=0,F=2,+$ and
A$^1\Pi,v=6,J=1,F=4,-$ $\leftarrow$ a$^3\Pi_0,v=5,J=0,F=3,+$ transitions, marked in red in Figure~\ref{fig:hyperfine2}, yield a lifetime of 2.1(3)~ns for the upper state hyperfine levels.
Since the values of $f_{\text{b}5}$ of both these levels differ by only 0.002, the lifetimes of these levels can be assumed to be identical and are shown as a single data point in Figure~\ref{fig:LifetimeSummary}.

%Cutted Voigt Fits
The transitions shown on the high frequency side in Figure~\ref{fig:hyperfine1} reach levels that have around 50~\% triplet character and the lifetimes of these levels are around twice as large as $\tau_{\text{A}6}$.
Lorentzian profiles convoluted with an adapted Doppler profile 
(70~MHz wide Doppler multiplied by a 20~MHz wide block function)
are used to fit the spectral lineshapes of the  two most isolated lines in the Doppler reduced spectrum.
These spectral lines belong to the 
A$^1\Pi, v=6$ // b$^3\Sigma^+, v=5, J=1, F=3, +$ $\leftarrow$ a$^3\Pi_1,v=5,J=1,F=2,-$ and
A$^1\Pi, v=6$ // b$^3\Sigma^+, v=5, J=1, F=2, +$ $\leftarrow$ a$^3\Pi_1,v=5,J=1,F=2,-$ transitions, marked in red in Figure~\ref{fig:hyperfine1}.
The thus determined lifetimes are 3.9(5)~ns and 4.2(5)~ns, respectively.

% time delayed ionization
The longest lifetimes are expected for the two hyperfine levels that belong to the $N=1$, $J=0$ level of the b$^3\Sigma^+, v=5$ state, as there is no $J=0$ level in the A$^1\Pi, v=6$ state to interact with. As $J$ is a good, but not a perfect quantum number, however, these hyperfine levels are calculated to possess a fraction of about 0.01 of singlet character, already sufficient to reduce their lifetimes relative to $\tau_{\text{b}5}$ by about a factor of two, i.e. down to about 100~ns. Such lifetimes can be conveniently measured via two-color, time-delayed ionization, using lasers with ns duration pulses. It is difficult to exclusively excite the $J=0$ level with a pulsed laser, as there are many hyperfine levels belonging to the same $N$ nearby. Instead, the lifetime of the hyperfine levels belonging to $N=0$ of the b$^3\Sigma^+, v=5$ state is measured, shown as the green curve in Figure~\ref{fig:expdecay}. Using time-delayed ionization, the lifetime for these levels is found as 44(3)~ns. Also indicated in Figure~\ref{fig:expdecay} is a triple resonance scheme that is used to selectively excite to the $F=6$ level of $N=2$, $J=3$ in the b$^3\Sigma^+, v=5$ state. This is one of the other levels predicted to have a large triplet character, and the lifetime of that level is found as 80(5)~ns.

%Lamb dip
Lifetimes in the intermediate, $5 -40$~ns range can be measured via spectral line broadening when Doppler-free techniques are employed. We perform Lamb dip measurements in the molecular beam by retro-reflecting the cw laser beam and recording the time-integrated LIF intensity generated by both the incoming and the reflected beam. The spectrally isolated transition from the $F=3$, $J=0$, $e$ level in the a$^3\Pi_0, v=5$ state to the $F=4$, $J=1$, $N=1$, $e$ level of the 
A$^1\Pi, v=6$ // b$^3\Sigma^+, v=5$ system at around 17286.316 cm$^{-1}$, the line marked in red in Figure~\ref{fig:hyperfine2}, is used for this. The calculated peak absorption cross section for this transition is equal to $\sigma_0 = 7.6 \cdot 10^{-12}$ cm$^2$. The $F=4$ level in the b$^3\Sigma^+, v=5$ state has a calculated fraction of triplet character of 0.899, and a calculated lifetime of $\tau(F=4)$ = 18.5(3.1)~ns. The error bar of this lifetime is calculated assuming an uncertainty of 10~\% of the lifetimes of both $\tau_{\text{A}6}$ and $\tau_{\text{b}5}$.

\begin{comment}
The error bar of this lifetime is solely determined by the precision with which the spin-orbit interaction parameter $\xi_{6,5}$ is known, assuming that the lifetimes $\tau_{\text{A}6}$ and $\tau_{\text{b}5}$ of the unperturbed states as given above are exact.
\end{comment}

In the experiment, we use about 3.5 mW of laser power at around 578 nm in a 
0.4 cm $\times$ 1.0 cm area, thus we have a photon flux of $\Phi \approx 2.5 \cdot 10^{16}$ cm$^{-2}$ s$^{-1}$ in both the incoming and the retro-reflected beam. The interaction time $T$ is well defined by the 4~mm long interaction region through which the molecules pass to be $T \approx 5.3$~$\mu$s. The product of the photon flux and the interaction time is thus given as $\Phi T \approx 1.3 \cdot 10^{11}$~cm$^{-2}$. The resulting LIF signal as a function of the laser frequency is shown in Figure~\ref{fig:Lamb-dip}. The observed spectral line is well described by a Voigt profile (dashed red curve), with a Gaussian contribution due to Doppler broadening of about 70 MHz, and with a Lorentzian Lamb dip  at the center.
The inset shows the (inverted) Lamb dip data together with the Lorentzian fit (blue curve).
The fit reveals the relative depth at the center frequency $\nu_0$ to be
$D(\nu_0)=0.26$(3), 
and the full width at half maximum to be $\gamma_\textrm{s} = 12.0$(2.4)~MHz. 
The error bars are the statistical errors from the fit. 

In the Appendix, we derive how the values for the cross section and the lifetime can be deduced from these two experimental observables in the limiting case of a fully open two-level system. Using Equations~(\ref{eq:sigmaA}) and (\ref{eq:tauA}), we find that $\sigma_0 = 5.3\text{(2.0)} \cdot 10^{-12}$ cm$^2$ and
$\tau = 18.7$(3.7)~ns.
The uncertainty in the experimental value for $\sigma_0$ is an estimate, based on the error bar in the value of $D(\nu_0)$ and the inhomogeneity of the laser intensity distribution within the rectangular area. The uncertainty in the value of $\tau$ is solely determined by the statistical uncertainty of the fit.
Both the experimentally determined value for the cross section and the value for the lifetime agree very well with the calculated values, contributing to an overall consistent picture of the singlet-triplet perturbation. 

\section{\label{sec:discussion}Discussion and Outlook
}

\begin{figure}[b]
		\centering
		\includegraphics[width=0.48\textwidth,clip]{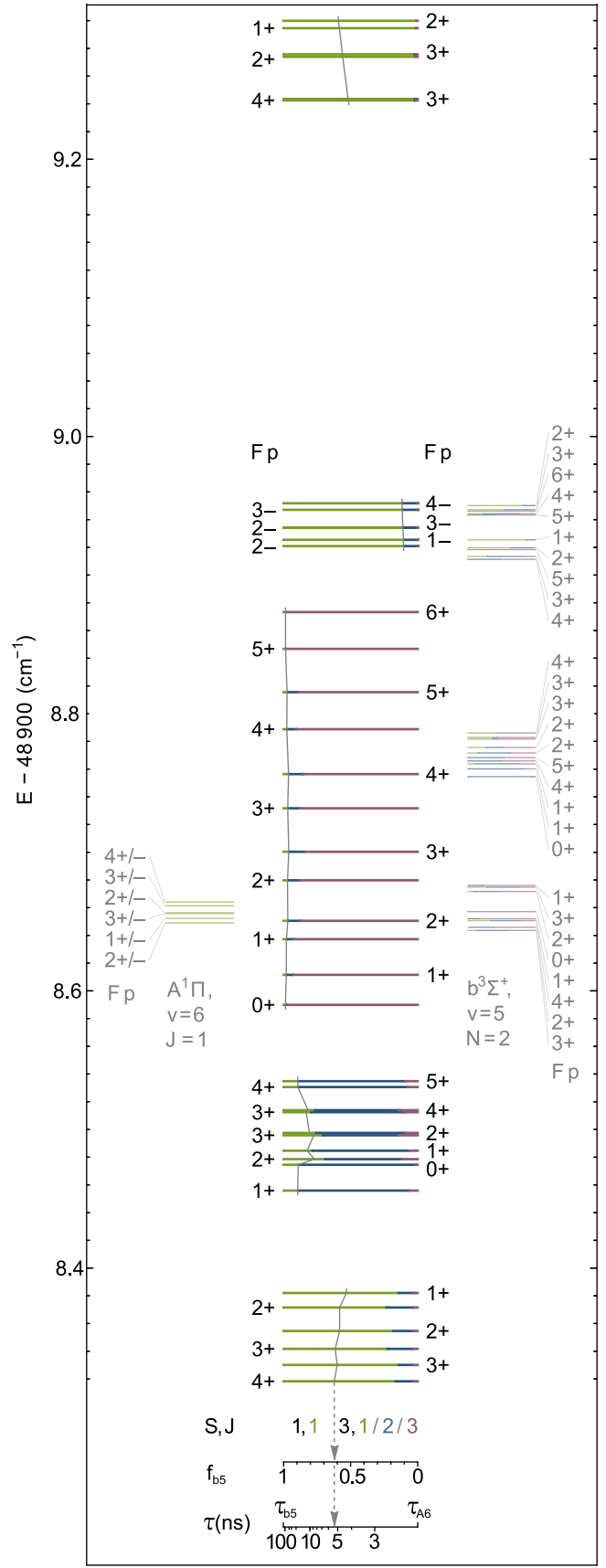}
	\caption{\label{fig:Fpert} Hyperfine levels of the perturbed A$^1\Pi, v=6, J=1$ and b$^3\Sigma^+, v=5, N=2$ system.}
\end{figure}
\FloatBarrier

The spectroscopic study of the perturbed A$^1\Pi, v=6$ // b$^3\Sigma^+, v=5$ system of AlF presented here has resulted in an accurate set of fine  and hyperfine structure parameters of the coupled states as well as in an accurate determination of their spin-orbit interaction parameter. A pictorial presentation of the main result of this study, showing the most relevant portion of the energy level diagram, is given in Figure~\ref{fig:Fpert}. A total of 34 positive parity and 6 negative parity hyperfine levels with a strongly mixed singlet-triplet character are located in an interval of about 1~cm$^{-1}$. On the left hand side in this Figure, the simulated hyperfine structure of the A$^1\Pi, v=6, J=1$ level is shown in the absence of spin-orbit coupling. On the right hand side, the same is shown for the b$^3\Sigma^+, v=5, N=2$ level. The actual energy levels, including the spin-orbit coupling, are depicted in the center. The $F$ quantum number and the parity of all the levels are given. The $J$ quantum number is less well defined and the $J$-fraction is indicated via color coding: green for $J=1$, blue for $J=2$ and violet for $J=3$. In the unperturbed A$^1\Pi, v=6$ state, $J$ is perfectly defined, whereas in the unperturbed $N=2$ level of the b$^3\Sigma^+, v=5$ state $J$ is not a useful quantum number. For the perturbed levels, the $J$ quantum number is seen to be quite well defined, with the fraction of a particular $J$ being always more than 50~\% and in most cases more than 80~\%. 

The gray wiggly lines crossing the energy levels indicate the fraction of singlet (left) and triplet (right) character. This fraction correlates directly with the lifetime of the levels, that range for these levels from 2.3~ns to 100~ns. It is seen in Figure~\ref{fig:LifetimeSummary} that the measured lifetimes agree very well with these predicted values, confirming the overall model.

Most of these levels are clearly separated from each other and can spectrally be selectively addressed. All these levels have a well defined and low value of $F$ ($0 \leq F \leq 6$), a well defined parity and, as a result of the perturbation, a reasonably well defined value of $J$ (when $0 \leq J \leq 3$). The combination of these properties makes these levels ideally suited as doorway states between the singlet and triplet manifold of electronic states in AlF.

\begin{acknowledgments}
We thank Marco De~Pas, Uwe Hoppe, Sebastian Kray and Klaus-Peter Vogelgesang for excellent technical support. N.~W. acknowledges support by the International Max Planck Research School for Elementary Processes in Physical Chemistry. This project has received funding from the European Research Council (ERC) under the European Union’s Horizon 2020 research and innovation programme (CoMoFun, Grant Agreement No. 949119).
\end{acknowledgments}

\section{Data Availability Statement}
The data that support the findings of this study are available from the corresponding author upon reasonable request.

\section*{Author Contributions}
N.~W. carried out the data acquisition, analyzed and visualized the data and wrote the manuscript. 
J.~S, S.~T and H.C.S. maintained the ring dye laser system,
B.~S. analyzed the data,
G.~M. supervised and conceptualized the project and edited the manuscript.

\appendix
\section{\label{sec:level1}Lamb dip experiments}  

In a Lamb dip experiment, a dip with a Lorentzian profile is observed, centered on top of a Doppler broadened line. The depth of the dip and the width of the Lorentzian profile contain information on the peak absorption cross section of the transition and on the lifetime of the upper level. The functional dependence of the depth and width of the Lamb dip on the cross section and lifetime can be found in textbooks for a fully closed two-level system.\cite{Demtroeder} Here the equations for a fully open two-level system are derived.

\subsection{Rate equation modelling of a general two-level system}

We start with a certain number of molecules \mbox{$N_a(t=0) = N^0_a$} at time $t=0$ in level $a$. This lower level is connected with a laser to an upper level $b$, that is originally empty, i.e.\ \mbox{$N_b(t=0) = 0$}. The energy separation between the levels is given by \mbox{$E_b - E_a$ = $h \, \nu_0$}, with the Planck's constant  $h$. Resonant emission from the upper level to the lower level occurs with rate $\Gamma_{\text{r}}$ (in s$^{-1}$). The total loss rate from the upper level, including $\Gamma_{\text{r}}$, is given by $\Gamma$ (in s$^{-1}$), which can be expressed as \mbox{$\Gamma = 1/\tau$}, where $\tau$ is the lifetime of the upper level. In the absence of any other broadening mechanisms, the frequency dependent absorption cross section $\sigma(\nu)$ (in cm$^2$) for the transition from level $a$ to level $b$ has a Lorentzian lineshape with a full width of half maximum given by $\gamma$ and can be written as
\begin{equation}
\sigma(\nu) = \sigma_0 \, \frac{(\gamma/2)^2}{(\nu-\nu_0)^2 + (\gamma/2)^2} \equiv \sigma_0 \, \mathcal{L}(\nu_0,\gamma)
\end{equation}
where $\sigma_0$ is the peak absorption cross section and $\mathcal{L}(\nu_0,\gamma)$ is a normalized Lorentzian profile (peak intensity normalized to one) centered around frequency $\nu_0$ and with a full width at half maximum $\gamma = \Gamma/(2 \pi) = 1/ (2 \pi \tau)$. When this two-level system interacts with a laser at frequency $\nu$ and with flux $\Phi$ (photons cm$^{-2}$ s$^{-1}$), the rate equations for this system can be written as:
\begin{equation}
\left( \begin{array} {c}
\frac{\textrm{d}N_a}{\textrm{d}t}\\[0.1cm]
\frac{\textrm{d}N_b}{\textrm{d}t}\\
\end{array} \right ) =
\left ( \begin{array}{rr}
- \sigma(\nu) \Phi & \sigma(\nu) \Phi + \Gamma_{\text{r}}\\[0.1cm]
\sigma(\nu) \Phi & - \sigma(\nu) \Phi - \Gamma \\[0.1cm]
\end{array} \right )
\left ( \begin{array} {c}
N_a \\[0.1cm]
N_b \\
\end{array} \right )
\end{equation}
The Eigenvalues $\lambda_{\pm}$ of this $2 \times 2$ matrix are given by:
\begin{equation}
\begin{aligned}
\lambda_{\pm} = - (\sigma(\nu)& \Phi +\Gamma/2 ) \\ \pm &\sqrt{(\sigma(\nu) \Phi)^2 +(\Gamma/2)^2 + \sigma(\nu) \Phi \Gamma_{\text{r}}}
\end{aligned}
\end{equation} 
The number of molecules in level $b$ at a given time can be expressed as:
\begin{equation}
N_b(t) = N^0_a  \frac{(\sigma(\nu) \Phi + \lambda_-)(\sigma(\nu) \Phi + \lambda_+)}{(\sigma(\nu) \Phi + \Gamma_{\text{r}}) (\lambda_- - \lambda_+)} \left ( e^{\lambda_+ t} - e^{\lambda_- t} \right )
\end{equation}
The number of photons, $n_{\textrm{fl}}(\nu,T)$, that are emitted via spontaneous fluorescence from level $b$ when the laser is interacting with the molecules for a time duration $T$ is given by
\begin{equation}
n_{\textrm{fl}}(\nu,T) \propto \int_{0}^{T} \Gamma N_b(t) \,\textrm{d}\,t.
\end{equation}

\subsection{\label{sec:level2}Lamb dips in an open two-level system}

In an open two-level system, the relation $\Gamma_{\text{r}} \ll \Gamma$ holds. In the situation that $\sigma_0 \Phi \ll \Gamma$ we can approximate $\lambda_+ \approx -\sigma(\nu) \Phi$ and $\lambda_- \approx -\Gamma$. The number of fluorescence photons, $n_{\textrm{fl}}(\nu,T)$, is given by
\begin{equation}
\label{eq:nfl0}
n_{\textrm{fl}}(\nu,T) \propto N^0_a \sigma(\nu) \Phi  \int_{0}^{T}  \left ( e^{- \sigma(\nu) \Phi t} - e^{- \Gamma t} \right ) \,\textrm{d}\,t.
\end{equation}
When the relaxation is fast, the second term under the integral can be neglected, and we find
\begin{equation}
\label{eq:nfl}
n_{\textrm{fl}}(\nu,T) \propto N^0_a ( 1 - e^{- \sigma(\nu) \Phi T} ) = N^0_a (1 - e^{- \sigma_0 \Phi T \mathcal{L}(\nu_0, \gamma)}). 
\end{equation}
Using that we can write
\begin{equation}
\frac{1 - e^{- \sigma(\nu) \Phi T}}{1-e^{-\sigma_0 \Phi T}} = \frac{e^{\sigma_0 \Phi T/2}(e^{\sigma(\nu) \Phi T/2} - e^{- \sigma(\nu) \Phi T/2})}{e^{\sigma(\nu) \Phi T/2}(e^{\sigma_0 \Phi T/2} - e^{- \sigma_0 \Phi T/2})}. 
\end{equation}
It is seen from a series expansion, including all terms linear and quadratic in $\sigma(\nu) \Phi T$ and $\sigma_0 \Phi T$, that the expression for $n_{\textrm{fl}}(\nu,T)$ can be approximated by
\begin{equation}
\label{eq:a9}
n_{\textrm{fl}}(\nu,T) \propto  N^0_a (1 - e^{- \sigma_0 \Phi T}) \mathcal{L}(\nu_0,\gamma_\textrm{fl})
\end{equation}
with 
\begin{equation}
\gamma_\textrm{fl} = \gamma \sqrt{ 1 + \frac{1}{2}\sigma_0 \Phi T + \frac{1}{8} (\sigma_0 \Phi T)^2 }.
\end{equation}
This is a broadened Lorentzian profile that is a very good approximation for $n_{\textrm{fl}}(\nu,T)$ given by Equation~(\ref{eq:nfl}) as long as $\sigma_0 \Phi T < 2$.

In a typical Lamb dip experiment, there is inhomogeneous broadening as well as saturation broadening of the Lorentzian line-profile, and the absorption line is described by a convolution of these profiles. When we retroreflect a mononchromatic laser beam with flux $\Phi$, i.e.\ when we have two counter-propagating laser beams, each with flux $\Phi$, the total number of spontaneous fluorescence photons that is emitted when the laser beams interact with different velocity-groups of molecules, is given by two times the value of $n_{\textrm{fl}}(\nu,T)$ given in Equation~(\ref{eq:a9}). When both laser beams interact with the same velocity-group of molecules, then this results in a total number of spontaneous fluorescence photons given by the value of $n_{\textrm{fl}}(\nu_0,T)$ for the doubled laser flux, $2 \Phi$, which is lower. This causes the appearance of the Lamb dip. In an open two-level system, the relative depth, $D(\nu_0)$, of the Lamb dip is thus given by 
\begin{equation}
D(\nu_0) = 1 - \frac{(1-e^{-2 \sigma_0 \Phi T})}{2(1-e^{- \sigma_0 \Phi T})} = \frac{1}{2} ( 1 - e^{- \sigma_0 \Phi T} )
\end{equation}
The shape is Lorentzian, saturation broadened due to the excitation rate $2 \sigma_0  \Phi$, and the full spectral profile of the Lamb dip is given by 
\begin{equation}
D(\nu) = \frac{(1 - e^{- 2 \sigma_0 \Phi T \mathcal{L}(\nu_0, \gamma)})}{2(1 + e^{- \sigma_0  \Phi T})}  \approx D(\nu_0) \mathcal{L}(\nu_0,\gamma_\textrm{s})
\end{equation}
with 
\begin{equation}
\gamma_\textrm{s} = \gamma \sqrt{ 1 + \sigma_0 \Phi T + \frac{1}{2} (\sigma_0 \Phi T)^2 }.
\end{equation}

The only experimental observables in a Lamb dip experiment are the relative depth of the Lamb dip ($0 < D(\nu_0) < 1/2$) and its full width at half maximum $\gamma_\textrm{s}$. In an open two-level system, the absorption cross section $\sigma_0$ is given by
\begin{equation}
\label{eq:sigmaA}
    \sigma_0 = \frac{1}{\Phi T} \textrm{ln} \left(\frac{1}{1- 2 D(\nu_0)}\right)  
  \,\, \stackrel{D(\nu_0) \ll1}{\approx}   \,\,
\frac{2 D(\nu_0)}{ \Phi T}
\end{equation}
and the lifetime $\tau$ is given by
\begin{equation}
\label{eq:tauA}
\begin{aligned}
\tau = &\frac{1}{2 \pi \gamma_\textrm{s}} \sqrt{ 1+ \textrm{ln} \left(\frac{1}{1-2D(\nu_0)}\right) + \frac{1}{2} \left(
\textrm{ln} \left(\frac{1}{1-2D(\nu_0)}\right)\right)^2 } \\[0.3cm]
 \,\, &\stackrel{D(\nu_0) \ll1}{\approx}   \,\,
\frac{1+ D(\nu_0)}{2 \pi \gamma_\textrm{s}}.
\end{aligned}
\end{equation}

\begin{comment}
When $D(\nu_0) \ll 1/2$, the peak absorption cross section is approximately given by $\sigma_0 \approx 2 D(\nu_0) / \Phi T$ and the lifetime by $\tau \approx (1+D(\nu_0))/2 \pi W$.
\end{comment}

\bibliography{aipsamp}% Produces the bibliography via BibTeX.

\end{document}